\documentclass[11pt]{article}
\usepackage{graphicx}
\usepackage{amssymb}
\usepackage{amsmath}    
\usepackage{amsthm}    
\usepackage{bm}         
\usepackage{upgreek}    
\usepackage{mathrsfs}  
\usepackage{extarrows}  
\usepackage{lscape}
\usepackage{threeparttable}
\usepackage{dcolumn}
\usepackage{multirow}
\usepackage{booktabs}
\usepackage{float}
\usepackage[top=1in, bottom=1.25in, left=1in, right=1in]{geometry}
\newcolumntype{d}[1]{D{.}{.}{#1}}
\RequirePackage{algorithm}%
\RequirePackage{algorithmicx}%
\RequirePackage{algpseudocode}%
\usepackage{caption}
\usepackage{subcaption}
\usepackage{comment}
\usepackage{enumitem}

\newtheorem{Example}{Example}[section]   

\newtheorem{Theorem}{Theorem}
\newtheorem*{Theorem*}{Theorem}        
\newtheorem{Lemma}{Lemma}
\newtheorem{Remark}{Remark}

\newtheorem*{Remark*}{Remark*}

\renewcommand{\footnotesize}{\fontsize{9pt}{11pt}\selectfont}

\makeatletter
\renewcommand\section{\@startsection{section}{1}{\z@}%
                                  {-3.5ex \@plus -1ex \@minus -.2ex}%
                                  {2.3ex \@plus.2ex}%
                                  {\normalfont\large\bfseries}}
\makeatother

\allowdisplaybreaks

\begin{document}

\title{\textbf{\large{Calculating the Upper Bound of Gini Coefficient Using Grouped Data and a Case Study of China}}}

\date{}
\renewcommand\thefootnote{}
\footnotetext{Pixu Shi is Assistant Professor at Biostatistics \& Bioinformatics, Duke University (E-mail: pixu.shi@duke.edu). Anru R Zhang is Eugene Anson Stead, Jr. M.D. Associate Professor at the Department of Biostatistics \& Bioinformatics and Department of Computer Science, Duke University (E-mail: anru.zhang@duke.edu).
The majority of this work was done while both authors were undergraduate students at Peking University. The authors were grateful to Professors Jiading Chen and Xiangzhong Fang for their supervision and to Professor Yanxia Ren for her valuable suggestions.}
\author{\normalsize{Pixu Shi~ and ~Anru R. Zhang}}
\maketitle
\begin{center}
{\it In memory of Professor Jiading Chen}
\end{center}
\medskip
\begin{abstract}
Determining an upper bound, particularly the optimal upper bound of the Gini coefficient when dealing with grouped data without specified income brackets, remains an important and open question. In this paper, we introduce an efficient algorithm to calculate the exact optimal upper bound of the Gini coefficient with provable guarantees. To exemplify these methods, we also offer computed results for the Gini coefficients of urban and rural China spanning the years 2003 to 2008.
\\
\\
\vspace*{.3in}
\noindent\textsc{KEYWORDS}: {Gini coefficient, Grouped data of income, Upper bound, China}
\\
\vspace*{.3in}
\noindent\textsc{MSC (2020)}: {91B82, 91B64}
\end{abstract}

\section{Introduction}\label{sec:intro}

The Gini coefficient is a crucial and popular index used to measure wealth disparity in specific countries or regions \cite{dorfman1979formula,gini1936measure}. Accurate calculation of the Gini coefficient typically requires complete income distribution data of a population. However, such detailed information is often unavailable. Many countries and regions release data in aggregated groupings rather than individual details. A notable example is the data spanning from 2003 to 2008, as presented in the {\it China Statistical Yearbook} (see Table~\ref{tab:data1} and Table~\ref{tab:data2}). In this dataset, respondents are categorized into distinct groups based on ascending per capita annual income, with eight groups for urban China and five for rural China. The data provides average per capita annual income and corresponding population proportions for each group but does not indicate income brackets. This provides partial information about the income distribution of the population without disclosing details at the individual level. Since the 1970s, effectively estimating the Gini coefficient from this type of data has been a significant challenge.

To describe the mathematical formulation of this challenge, we introduce the non-negative random variable $Y$ to represent the per capita annual income within a specific country or region. Let $F(x)=\mathbb{P}(Y\leqslant x)$ represent the distribution function of $Y$. We also define $\mu=\mathbb{E}Y\in(0,+\infty)$ and $F^{-1}(p)=\inf\{x:F(x)>p\}, p\in[0,1)$. With these definitions in place, the {\it Lorenz function} and {\it Gini coefficient} are denoted as $L(p)$ and $G$, respectively:
\begin{equation}\label{eq:lorenz}
L(p)=\frac{1}{\mu}\int_0^pF^{-1}(u)du\quad (0\leqslant p\leqslant1),
\end{equation}
\begin{equation}\label{eq:Gini}
G=1-2\int_0^1L(p)dp.
\end{equation}
We can see the actual value of the Gini coefficient $G$ relies on the distribution distribution function $F$, which is often unknown in practice.

Since many countries and regions publish the {\it grouped data} and {\it income brackets} for per capita annual income defined as follows, we define $(a_{i-1},a_i]$ as the income bracket of Group $i$ for $i=1,\dots,k+1$, and $[a_0,a_1]$ for Group 1, with 
$0=a_0<a_1<\cdots<a_k<a_{k+1}=\infty$. Define
\begin{equation}\label{eq:p}
p_0=0, \quad p_i=F(a_i),\quad i=1,\ldots,k+1,
\end{equation}
\begin{equation}\label{eq:mu}
\mu_1=\mathbb{E}(Y|Y\leqslant a_1), \quad\mu_i=\mathbb{E}(Y|a_{i-1}<Y\leqslant a_i),\quad i=2,\ldots,k+1.
\end{equation}
Here, $\mu_i$ represents the theoretical average annual income of Group $i$, while $p_i-p_{i-1}$ is its theoretical population proportion of Group $i$ that is generally nonzero. When the collection of data is done through proper survey sampling with a large number of respondents, as in most real-world scenarios, we can confidently approximate $p_i-p_{i-1}$ and $\mu_i$ with their empirical equivalents.

The goal of our paper is to establish bounds for the Gini index by leveraging $\mu_i$ and $p_i-p_{i-1}$, the limited published information on income groups, when the full income distribution $F(x)$ is unknown. Such bounds will provide valuable information on wealth disparity for policymakers and researchers without the need to acquire fully disclosed income data.

\subsection{Results in the Literature}\label{sec:related-result}
In the literature, various estimators of $G$ were proposed by incorporating assumptions about either $F(x)$ or $L(p)$; however, these estimators may heavily rely on the chosen assumptions \cite{cheong2002empirical,ogwang2006upper,schader1994fitting}. Alternatively, \cite{gastwirth1972estimation,mehran1975bounds,silber1990new} pursued the identification of upper and lower bounds for the Gini coefficient $G$ without imposing any assumptions on $F(x)$ or $L(p)$. Notably, if these upper and lower bounds are in close proximity, they shed light on the characteristics of  $G$.

According to \eqref{eq:lorenz}, \eqref{eq:p}, and \eqref{eq:mu},
\begin{equation}
\mu=\mathbb{E}Y=\sum_{i=1}^{k+1}(p_i-p_{i-1})\mu_i,
\end{equation}
\begin{equation}\label{eq:Lp}
L(p_i)=\sum_{j=1}^{i}(p_j-p_{j-1})\frac{\mu_j}{\mu}.
\end{equation}
Thus, the values of $\{\mu, p_i, L(p_i)\}$ can be calculated from $\{p_i, \mu_i\}$ and vice versa. Gastwirth \cite{gastwirth1972estimation} first provided the optimal lower bound $GL$ based on the values of $\{\mu, p_i,L(p_i)\}$:
\begin{equation}\label{eq:GL}
GL=1-\sum_{i=1}^{k+1}(p_i-p_{i-1})(L(p_i)+L(p_{i-1})).
\end{equation}
Utilizing $\{p_i,\mu_i\}$ along with the known values of $\{a_i\}$, \cite{gastwirth1972estimation} also provided an upper bound $GU$ as
\begin{equation}\label{eq:GUbracket}
GU=GL+D,\quad D\triangleq\frac{1}{\mu}\left\{\sum_{i=1}^{k}(p_i-p_{i-1})^2\frac{(a_i-\mu_i)(\mu_i-a_{i-1})}{a_i-a_{i-1}}+(1-p_k)^2(\mu_{k+1}-a_k)\right\}.
\end{equation}
However, \eqref{eq:GUbracket} cannot be applied when $\{a_i\}$ is absent, and in this scenario, Mehran \cite{mehran1975bounds} introduced a optimal upper bound $GU$ through a geometric approach, based on the assumption of the Lorenz function's differentiability across its entire domain:
\begin{equation}\label{eq:Mehran}
GU=GL+\Delta^*, \quad\Delta^*\triangleq\underset{\beta_i^*\in[\beta_i,\beta_{i+1}], i=1,\ldots,k}{\sup}\Delta(\beta_1^*,\ldots,\beta_k^*),
\end{equation}
where
\begin{equation}\label{eq:aim}
\begin{split}
\Delta(\beta_1^*,\ldots,\beta_k^*) &=(p_1-p_0)^2(\beta_1^*-\beta_1)\frac{\beta_1}{\beta_1^*}+(p_{k+1}-p_{k})^2(\beta_{k+1}-\beta_{k}^*)\\
&\quad +\sum_{i=2}^{k}(p_{i}-p_{i-1})^2\frac{(\beta_{i}^*-\beta_i)(\beta_i-\beta_{i-1}^*)}{\beta_{i}^*-\beta_{i-1}^*},\\
\end{split}
\end{equation}
\begin{equation}\label{eq:beta}
\beta_i=\frac{L(p_{i})-L(p_{i-1})}{p_{i}-p_{i-1}}=\frac{\mu_i}{\mu},\quad i=1,\ldots,k+1.
\end{equation}
In \eqref{eq:aim}, we set $0/0=0$ for convenience of notation. 
\begin{Remark}[A Geometric Interpretation of Mehran's Upper Bound $GU$ \eqref{eq:Mehran}--\eqref{eq:beta}]
In Figure \ref{fig:illustration}, we plot $(p_i, L(p_i))$ and connect them with black solid polylines. We also set $l_i$ as the straight line passing through $(p_i, L(p_i))$ with the slope of $\beta_i^*$ and draw them as purple solid polylines. Given this setup, $A_1,\ldots, A_k$ ($k=5$ in the context of Figure \ref{fig:illustration}) are the triangular areas surrounded by black and purple polylines, highlighted in Figure \ref{fig:illustration}. The following equations hold:
\begin{equation*}
    \begin{split}
        \text{Area}(A_1) = & \frac{1}{2}(p_1-p_0)^2(\beta_1^*-\beta_1)\frac{\beta_1}{\beta_1^*};\\
\text{Area}(A_i) = & \frac{1}{2}(p_{i}-p_{i-1})^2\frac{(\beta_{i}^*-\beta_i)(\beta_i-\beta_{i-1}^*)}{\beta_{i}^*-\beta_{i-1}^*}, i=2,\ldots, k-1;\\
    \text{Area}(A_k) = & \frac{1}{2}(p_{k+1}-p_{k})^2(\beta_{k+1}-\beta_{k}^*).
    \end{split}
\end{equation*}
Therefore, $\Delta/2$ is the area between the black solid polyline that connects $(p_i, L(p_i))$ with the purple polylines that intersect $l_i$ in sequence.
\begin{figure}
    \centering
    \includegraphics[width=10cm]{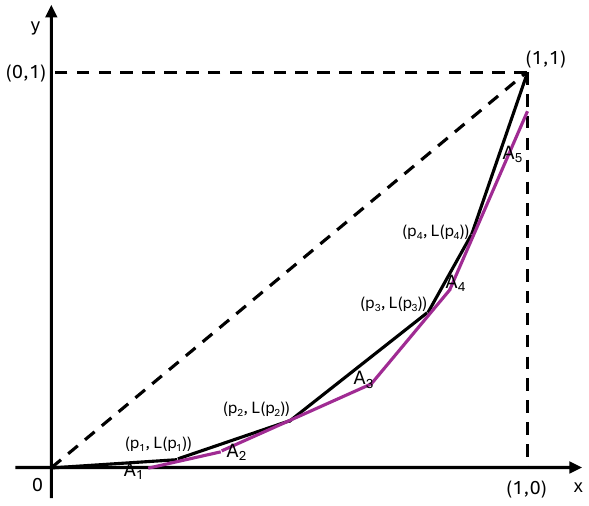}
    \caption{A pictorial illustration of Mehran's upper bound $GU$.}
    \label{fig:illustration}
\end{figure}
\end{Remark}

\cite{mehran1975bounds} claimed that $\Delta^*$ is reached at $\{\beta_i^*\}$:
\begin{equation}\label{eq:MehranBeta}
\beta_i^*=\left\{
              \begin{array}{ll}
                b_i^*\triangleq \frac{L(p_i)-c_i}{p_i-d_i}, & \hbox{$\beta_i\leqslant b_i^*\leqslant\beta_{i+1}$;} \\
                \beta_i, & \hbox{$b_i^*<\beta_i$;} \\
                \beta_{i+1}, & \hbox{$b_i^*>\beta_{i+1}$,}
              \end{array}
            \right.
\end{equation}
where $\{d_i\}$ and $\{c_i\}$ are recursively defined by
\begin{equation}
\begin{split}
&d_{k+1}=1,\quad d_i=2p_i-d_{i+1},\quad i=1,\ldots,k\\
&c_1=0,\quad c_i=2L(p_{i-1})-c_{i-1},\quad i=2,\ldots,k+1.\\
\end{split}
\end{equation}
However, our Example~\ref{ex:Mehran1} in Section~\ref{sec:example} serves as a counter-example to show this assertion is incorrect. Moreover, \eqref{eq:MehranBeta} cannot be applied when $p_i=d_i$ for some $i$. In the specific scenario where $p_i=\frac{i}{k+1}$, \cite{mehran1975bounds} has proposed an alternative method as follows:
\begin{equation}\label{eq:equally}
\beta_i^*=\beta_{k+1-2[\frac{1}{2}(k+2-i)]+1},\quad i=1,\ldots,k,
\end{equation}
where ``$[x]$'' represents the largest integer no greater than $x$. 
Example~\ref{ex:Mehran2} provides another counter-example to show this result is not correct either. \cite{silber1990new} wrote that $\Delta^*$ is reached at $\beta_i^*$ given by
\begin{equation}\label{eq:Silber}
\beta_i^* = \frac{L(p_{i+1})-L(p_{i-1})}{p_{i+1}-p_{i-1}}
\end{equation}
Our counter-example in Example \ref{ex:Silber} shows this is also incorrect.

Furthermore, \cite{cerone2007bounds,fuller1979estimation,giorgi1987general,murray1978extreme} have also explored the upper bounds for the Gini coefficient with the knowledge of income brackets ${a_i}$. On a related note, \cite{ogwang2003bounds,ogwang2004modification} established a connection between \eqref{eq:GUbracket} and \eqref{eq:Mehran}; \cite{langel2013variance} surveyed and discussed the methods for variance estimation of the Gini coefficient; \cite{modarres2006cautionary} considered estimating the standard error of the Gini coefficient; \cite{davidson2009reliable} studied the inference for the Gini coefficient; \cite{dai2013interval} studied the interval estimation of Gini coefficient. 

In summary, calculating the optimal upper bound of the Gini coefficient in the absence of income brackets remains an open problem. 

\subsection{Our Contributions and Paper Organizations}\label{sec:contributions}

In this paper, we aim to bridge this gap by introducing an efficient procedure to calculate the optimal upper bound of the Gini coefficient, in cases where income brackets are unavailable. In Section \ref{sec:minimum-upper-bound-Gini}, we establish the correctness of Formula~\eqref{eq:Mehran} (Theorem~\ref{th:traditional}) without any smoothness assumption on $L(p)$. In Section \ref{sec:new-method}, we devise an efficient method to calculate the optimal upper bound of the Gini coefficient based on grouped income data when specific income brackets are not provided. We present a fine and explicit-form upper bound, denoted as $\widetilde{GU}$, and demonstrate that, under certain conditions detailed in Section \ref{sec:upper-bound-gini}, $\widetilde{GU}$ aligns with the optimal upper bound. Expanding on $\widetilde{GU}$, we introduce an efficient recursive algorithm to calculate the exact optimal upper bound $GU$ in Section \ref{sec:algorithm}. We then apply the proposed algorithm in conjunction with the lower bound \eqref{eq:GL} to give a precise range of Gini coefficient for rural and urban China between 2003 and 2008 in Section \ref{sec:China}, utilizing data sourced from the {\em China Statistical Yearbook}. The counterexamples and additional proofs can be found in Section \ref{sec:example} and the Appendix, respectively.

\section{A General Formula for the Optimal Upper Bound}\label{sec:minimum-upper-bound-Gini}

In this section, we prove that $GU$ given by \eqref{eq:Mehran} \eqref{eq:aim} \eqref{eq:beta} is the optimal upper bound of the Gini coefficient in general. Our result aligns with the finding of Mehran \cite{mehran1975bounds} without any smoothness assumption.
\begin{Theorem}\label{th:traditional}
Recall $L(p)$ is a Lorenz function (defined in \eqref{eq:lorenz}), $(p_i,L(p_i))$ $(i=0,\ldots,k+1)$ are $k+2$ points (including $(0,0)$ and $(1,1)$) on $L(p)$. Then $GL +\Delta^*$ is the optimal upper bound of Gini coefficient, where $\Delta^*$ is given by \eqref{eq:Mehran} and \eqref{eq:aim}.
\end{Theorem}
\noindent\textbf{Proof of Theorem \ref{th:traditional}.} To prove Theorem~\ref{th:traditional}, we first present Lemma~\ref{lemma1}.
\begin{Lemma}\label{lemma1}
Suppose $f(x)$ is a continuous, convex, and nondecreasing function on $[a,b]$. Since $f(x)$ is convex, $f$ has both left and right derivatives. $\beta=\frac{f(b)-f(a)}{b-a}$, $u=f_+'(a)$, $v=f_-'(b)$, $h(x)=f(a)+\beta(x-a)$, where $f_+'(a)$ and $f_-'(b)$ are the right derivative of $f$ at $a$ and the left derivative of $f$ at $b$, respectively. Then:
\begin{enumerate}
  \item when $v<\infty$, $\int_a^b(h-f)dx\leqslant\frac{1}{2}(b-a)^2\frac{(v-\beta)(\beta-u)}{v-u}$;
  \item when $v=+\infty$, $\int_a^b(h-f)dx\leqslant\frac{1}{2}(b-a)^2(\beta-u)$.
\end{enumerate}
\end{Lemma}
The proof of Lemma \ref{lemma1} is provided in the Appendix. Next, we define $H(p)=\underset{i=1,\ldots,k+1}{\max}h_i(p)$, where
$$h_i(p)=\frac{L(p_i)-L(p_{i-1})}{p_i-p_{i-1}}(p-p_{i-1})+L(p_{i-1}),\quad i=1,\ldots,k+1.$$
Lorenz function is a continuous convex monotonic nondecreasing function on $[0,1]$ according to its definition \eqref{eq:lorenz}. Therefore, $H(p)$ is the piecewise linear function formed by intersecting $h_1(p),\ldots,h_{k+1}(p)$ in sequence. According to Lemma~\ref{lemma1},
$$\int_{p_{i-1}}^{p_i}(H(p)-L(p))dp\leqslant
\frac{1}{2}(p_i-p_{i-1})^2\frac{(L'_-(p_i)-\beta_i)(\beta_i-L'_+(p_{i-1}))}{L'_-(p_i)-L'_+(p_{i-1})},\quad i=1,\ldots,k+1.$$
The right-hand side of the inequality above is nondecreasing in $L'_-(p_{i})$, thus,
$$\int_{p_{i-1}}^{p_i}(H(p)-L(p))dp\leqslant
\frac{1}{2}(p_i-p_{i-1})^2\frac{(L'_+(p_i)-\beta_i)(\beta_i-L'_+(p_{i-1}))}{L'_+(p_i)-L'_+(p_{i-1})},\quad i=1,\ldots,k+1.$$
Adding them up we get:
\begin{equation}\label{eq:proof1}
\begin{split}
\int_0^1(H(p)-L(p))dp
& \leqslant \frac{1}{2}\sum_{i=1}^{k+1}(p_i-p_{i-1})^2\frac{(L'_+(p_i)-\beta_i)(\beta_i-L'_+(p_{i-1}))}{L'_+(p_i)-L'_+(p_{i-1})}\\
& \leqslant \sup_{\beta_i^*\in[\beta_i,\beta_{i+1}],i=0,\ldots,k+1}\frac{1}{2}\sum_{i=1}^{k+1}(p_i-p_{i-1})^2\frac{(\beta_i^*-\beta_i)(\beta_i-\beta_{i-1}^*)}{\beta_i^*-\beta_{i-1}^*}\\
&=\sup_{\beta_i^*\in[\beta_i,\beta_{i+1}],i=1,\ldots,k}\Big\{\frac{1}{2}\sum_{i=2}^{k}(p_i-p_{i-1})^2\frac{(\beta_i^*-\beta_i)(\beta_i-\beta_{i-1}^*)}{\beta_i^*-\beta_{i-1}^*}\\
&+\frac{1}{2}(p_1-p_0)^2(\beta_1^*-\beta_1)\frac{\beta_1}{\beta_1^*}+\frac{1}{2}(p_{k+1}-p_{k})^2(\beta_{k+1}-\beta_{k}^*)\Big\}\\
&= \frac{1}{2}\sup_{\beta_i^*\in[\beta_i,\beta_{i+1}],i=1,\ldots,k}\Delta = \frac{1}{2}\Delta^*.\\
\end{split}
\end{equation}
Suppose that the supremum in \eqref{eq:proof1} is reached at $\tilde{\beta_i^*},(i=1,\ldots,k)$. Define $\tilde{L}(p)=\underset{i=1,\ldots,k+1}{\max}\{\tilde{\beta_i^*}(p-p_i)+L(p_i)\}$.
Then $\tilde{L}(p)$ is a continuous convex nondecreasing function between $(0,0)$ and $(1,1)$, which could be the Lorenz Function corresponding to some distribution. From \eqref{eq:GL} we know that $1-2\int_0^1H(p)dp=GL$. Thus
\begin{equation}
G=1-2\int_0^1L(p)dp \leqslant 1-2\int_0^1H(p)dp +\Delta^* = GL +\Delta^*=1-2\int_0^1\tilde{L}(p)dp.
\end{equation}
Therefore, $GL +\Delta^*$ is the optimal upper bound of Gini coefficient $G$.
\qed

Theorem \ref{th:traditional} implies that the evaluation of the optimal upper bound of $G$ boils down to the following optimization:
\begin{equation}\label{eq:maximization}
\begin{split}
\max_{\beta_1^*, \ldots, \beta_k^*} \quad & \Delta(\beta_1^*,\ldots, \beta_k^*) = (p_1-p_0)^2(\beta_1^*-\beta_1)\frac{\beta_1}{\beta_1^*}+(p_{k+1}-p_{k})^2(\beta_{k+1}-\beta_{k}^*)\\
& \qquad\qquad\qquad\qquad  +\sum_{i=2}^{k}(p_{i}-p_{i-1})^2\frac{(\beta_{i}^*-\beta_i)(\beta_i-\beta_{i-1}^*)}{\beta_{i}^*-\beta_{i-1}^*},\\
& \text{where} \quad \beta_i=\frac{L(p_{i})-L(p_{i-1})}{p_{i}-p_{i-1}},\quad i=1,\ldots,k+1,\\
\end{split}
\end{equation}
\begin{equation}\label{ineq:constraint-beta}
    \text{subject to} \quad \beta_1\leqslant \beta_1^* \leqslant \beta_2 \leqslant \beta_2^* \leqslant \cdots \leqslant \beta_{k} \leqslant \beta_{k}^* \leqslant \beta_{k+1}.\qquad\qquad\qquad\qquad\quad
\end{equation}
\section{New Method to Calculate the Optimal Upper Bound}\label{sec:new-method}

In this section, we introduce a novel approach to determine the optimal upper bound of $GU$ by presenting an explicit solution to \eqref{eq:maximization}. To begin with, we establish a closed-form upper bound for the Gini coefficient, denoted as $\widetilde{GU}$, in Section \ref{sec:upper-bound-gini}. Under specific conditions, we can demonstrate that $\widetilde{GU}$ is the optimal upper bound. Furthermore, $\widetilde{GU}$ offers insights into a systematic sequential method for evaluating the actual optimal upper bound of the Gini coefficient, denoted as $GU$. Building upon this, in Section \ref{sec:algorithm}, we introduce a recursive method to compute the precise value of $GU$ with theoretical guarantees.

\subsection{A Fine Upper Bound $\widetilde{GU}$}\label{sec:upper-bound-gini}

We define $z_i$ $(i=1,\ldots,k+1)$ recursively:
\begin{equation}\label{eq:recursive}
z_{k+1}=1,\quad z_{k-i}=2p_{k-i}-z_{k-i+1},\quad i=0,\ldots, k-1.
\end{equation}
Then we define $\widetilde{GU}$ as:
\begin{equation}\label{eq:tildeGU}
\widetilde{GU}=GL+\beta_1z_1^{2}+\underset{i=1}{\overset{k}{\sum}}(\beta_{i+1}-\beta_i)(p_{i}-z_i)^2.
\end{equation}
Theorem \ref{th:unstrict1} below shows that $\widetilde{GU}$ is an upper bound for the Gini coefficient.
\begin{Theorem}[$\widetilde{GU}$ is an upper bound]\label{th:unstrict1}
The maximum of \eqref{eq:maximization} subject to the constraint \eqref{ineq:constraint-beta} satisfies
\begin{equation}\label{eq:unequal}
\Delta(\beta_1^*,\ldots, \beta_k^*) \leqslant \beta_1z_1^{2}+\sum_{i=1}^{k}(\beta_{i+1}-\beta_i)(p_{i}-z_i)^2.
\end{equation}
Accordingly, $\widetilde{GU}$ defined by \eqref{eq:tildeGU} is an upper bound of $G$, i.e., $GU \leq \widetilde{GU}$.
\end{Theorem}


Moreover, the upper bound $\widetilde{GU}$ developed in Theorem \ref{th:unstrict1} is the optimal upper bound under some further conditions, as presented in Theorem~\ref{th:unstrict2}.
\begin{Theorem}[$\widetilde{GU} = GU$ under further conditions]\label{th:unstrict2}
Define $w_i$ recursively for $i=1,\ldots, k$:
\begin{equation}\label{eq:recursive2}
w_{1}=0,\quad w_{i}=2L(p_{i-1})-w_{i-1},\quad i=2,\ldots, k+1.
\end{equation}
Define $B_i, i=1,\ldots,k$, as
\begin{equation}\label{eq:betastar}
B_i\triangleq
\left\{
  \begin{array}{ll}
    \frac{w_{i+1}-w_i}{z_{i+1}-z_i}, & \hbox{when $z_{i}\neq z_{i+1};$} \\
    \infty, & \hbox{when $z_i=z_{i+1}$ and $w_i\neq w_{i+1};$} \\
    \frac{1}{2}(\beta_i+\beta_{i+1}), & \hbox{when $z_i=z_{i+1}$ and $w_i=w_{i+1}.$}
  \end{array}
\right.
\end{equation}
If the following inequality holds:
\begin{equation}\label{eq:conditionB}
\beta_1<B_1<\beta_2<B_2<\cdots<\beta_k<B_k<\beta_{k+1},
\end{equation}
then $\widetilde{GU}$ equals the actual optimal upper bound $GU$:
\begin{equation}\label{eq:GUnotbest}
GU=\widetilde{GU}=GL+\beta_1z_1^{2}+\underset{i=1}{\overset{k}{\sum}}(\beta_{i+1}-\beta_i)(p_{i}-z_i)^2.\\
\end{equation}
\end{Theorem}

\subsection{A Recursive Algorithm to Calculate $GU$}\label{sec:algorithm}

While the previous section introduced $\widetilde{GU}$, which offers an explicit upper bound for the Gini coefficient when dealing with grouped data, $\widetilde{GU}$ may be strictly greater than the optimal upper bound if the conditions specified in \eqref{eq:conditionB} are not met. In this section, we further introduce a new method to evaluate the exact optimal upper bound for the Gini coefficient based on the grouped data. 

When $k+1=1$, there is only one group, and $\Delta^*=1$. Therefore, we will exclusively address situations where $k+1\geqslant2$. Theorems~\ref{th:unstrict2} and \ref{th:enumerate} collectively present an algorithm for calculating the optimal upper bound of the Gini coefficient. Before stating Theorem~\ref{th:enumerate}, which is the primary result of this section, we first introduce Lemma~\ref{th:possible} through Lemma~\ref{th:trans}, and their proofs are collected in the Appendix.

The next two lemmas elucidate the permissible values for $\{\beta_i^*\}_{i=1}^k$ when they attain the optimum in the optimization problem \eqref{eq:maximization}, subject to the constraint \eqref{ineq:constraint-beta}. These lemmas establish fundamental guidelines that aid in determining $GU$.
\begin{Lemma}[A Necessary Condition for Optimality of $\{\beta_i^*\}$]\label{th:possible}
Consider the optimization problem \eqref{eq:maximization}. Suppose $\Delta$ achieves its maximum value $\Delta^*$ at $\{\beta_i^*\}_{i=1}^k$. Then each $\beta_i^*$ can only be in one of the following three situations:
\begin{enumerate}
  \item $\beta_i^*=\beta_{i}$ or $\beta_i^*=\beta_{i+1}$.
  \item \begin{equation}\label{eq:bstar}
  \beta_i^*=\frac{\beta_{i+1}^*(p_i-p_{i-1})(\beta_{i}-\beta_{i-1}^*)+\beta_{i-1}^*(p_{i+1}-p_i)(\beta_{i+1}^*-\beta_{i+1})}{(p_i-p_{i-1})(\beta_{i}-\beta_{i-1}^*)+(p_{i+1}-p_i)(\beta_{i+1}^*-\beta_{i+1})} \triangleq b_i^* \in (\beta_i,\beta_{i+1}).
      \end{equation}
  \item $\Delta$ does not rely on $\beta_i^*$, i.e., $\beta_i^*$ could be any value in $[\beta_i,\beta_{i+1}]$.
\end{enumerate}
\end{Lemma}

\begin{Lemma}[A Necessary Condition for Optimality of $\{\beta_i^*\}$ when \eqref{eq:conditionB} fails]\label{th:notmiddle}
Consider the optimization problem \eqref{eq:maximization}. If \eqref{eq:conditionB} fails, then in order to ensure that $\Delta$ reaches the maximum value $\Delta^*$ at $\{\beta_i^*\}$ subject to the constraint $\beta_1\leqslant \beta_1^* \leqslant \beta_2 \leqslant \beta_2^* \leqslant \cdots \leqslant \beta_{k} \leqslant \beta_{k}^* \leqslant \beta_{k+1}$, there must exist a $j \in \{1,\ldots,k\}$ such that $\beta_j^*$ belongs to Situation 1 or 3 in Lemma~\ref{th:possible}.
\end{Lemma}


Lemma \ref{th:notmiddle} demonstrates that if \eqref{eq:conditionB} is not satisfied, implying that the refined upper bound $\widetilde{GU}$ may not be the optimal upper bound, we have either $\beta_j^* = \beta_j$ or $\beta_{j+1}$. Building upon this observation, the subsequent lemma demonstrates that, in such situations, the target optimization problem can be divided into two smaller-scale sub-problems: optimizing $\Delta^{(j,1)}$ and $\Delta^{(j,2)}$.
\begin{Lemma}[Decomposition of $\sup\Delta$ when $\beta_j^* = \beta_j$ or $\beta_{j+1}$]\label{th:decompose}
For $j=1,\ldots,k+1$, define $\Delta^{(j,1)}$ and $\Delta^{(j,2)}$ as follows:
\begin{equation}\label{eq:decompose1}
\Delta^{(j,1)}(\beta_1^*,\ldots,\beta_{j-2}^*)=
\left\{
  \begin{array}{ll}
    0, & \hbox{$j=1$,} \\
    \underset{i=1}{\overset{j-2}{\sum}}\Delta_i+\Delta_{j-1}|_{\beta_{j-1}^*=\beta_{j}}, & \hbox{$j=2,\ldots,k+1,$}
  \end{array}
\right.
\end{equation}
\begin{equation}\label{eq:decompose2}
\Delta^{(j,2)}(\beta_{j+1}^*,\ldots,\beta_{k}^*)=
\left\{
  \begin{array}{ll}
    \Delta_{j+1}|_{\beta_{j}^*=\beta_{j}}+\underset{i=j+2}{\overset{k+1}{\sum}}\Delta_i, & \hbox{$j=1,\ldots,k,$} \\
    0, & \hbox{$j=k+1.$}
  \end{array}
\right.
\end{equation}
Then,
\begin{equation}
\underset{\beta_i^*\in[\beta_i,\beta_{i+1}],i=1,\ldots,j-1,j+1,\ldots,k}{\sup}\Delta|_{\beta_j^*=\beta_{j}}
=\underset{\beta_i^*\in[\beta_i,\beta_{i+1}],i=1,\ldots,j-2}{\sup}\Delta^{(j,1)}
+\underset{\beta_i^*\in[\beta_i,\beta_{i+1}],i=j+1,\ldots,k}{\sup}\Delta^{(j,2)},
\end{equation}
and
\begin{equation}
\underset{\beta_i^*\in[\beta_i,\beta_{i+1}],i=1,\ldots,j-1,j+1,\ldots,k}{\sup}\Delta|_{\beta_j^*=\beta_{j+1}}
=\underset{\beta_i^*\in[\beta_i,\beta_{i+1}],i=1,\ldots,j-1}{\sup}\Delta^{(j+1,1)}
+\underset{\beta_i^*\in[\beta_i,\beta_{i+1}],i=j+2,\ldots,k}{\sup}\Delta^{(j+1,2)}.
\end{equation}
\end{Lemma}


Moreover, the following lemma gives expressions for $\Delta^{(j,1)}$ and $\Delta^{(j,2)}$ in a form of $\Delta^{'(j,1)}$ and $\Delta^{'(j,2)}$. It is worth noting that $\Delta^{'(j,1)}$ and $\Delta^{'(j,2)}$ follow the same format as \eqref{eq:aim}, albeit with a reduced scale. The proof of Lemma~\ref{th:trans} logically proceeds from the preceding theorem and is omitted herein.
\begin{Lemma}[An expression of $\Delta^{'(j,1)}$ and $\Delta^{'(j,2)}$]\label{th:trans}
~
\begin{enumerate}
\item
Transform $(p_i,L(p_i))$ $(i=0,\ldots,j-1,j\geqslant2)$ and $\beta_i=\frac{L(p_i)-L(p_{i-1})}{p_i-p_{i-1}}$ $(i=1,\ldots,j-1,j\geqslant2)$ to $(p_i',L(p_i)')$ and $\beta_i'$ via the following affine transformation:
\begin{equation}
    \left\{
      \begin{array}{ll}
        x'=a_{j1}x+b_{j1}y,&\hbox{$a_{j1}=\frac{L(p_{j})-L(p_{j-1})}{p_{j-1}L(p_{j})-p_{j}L(p_{j-1})},b_{j1}=\frac{1-a_{j1}p_{j-1}}{L(p_{j-1})}$;}\\
        y'=c_{j1}x+d_{j1}y, &\hbox{$c_{j1}=0,d_{j1}=\frac{1}{L(p_{j-1})}$.}
      \end{array}
    \right.
    \end{equation}
Set $k'=j-2$. Then
\begin{equation}\label{eq:Deltastar1}
\underset{\beta_i^*\in[\beta_i,\beta_{i+1}],i=1,\ldots,j-2}{\sup}\Delta^{(j,1)}=
\begin{vmatrix}
      a_{j1} & b_{j1} \\
      c_{j1} & d_{j1} \\
      \end{vmatrix}^{-1}
\underset{\beta_i^*\in[\beta_i',\beta_{i+1}'],i=1,\ldots,j-2}{\sup}\Delta^{'(j,1)}
\triangleq\begin{vmatrix}
      a_{j1} & b_{j1} \\
      c_{j1} & d_{j1} \\
      \end{vmatrix}^{-1}(\Delta^{'(j,1)})^*,
\end{equation}
where
\begin{equation}
\begin{split}
\Delta^{'(j,1)}=&(p_1'-p_0')^2(\beta_1^*-\beta_1')\frac{\beta_1'}{\beta_1^*}
+(p_{k'+1}'-p_{k'}')^2(\beta_{k'+1}'-\beta_{k'}^*)\\
&+\sum_{i=2}^{k'}(p_{i}'-p_{i-1}')^2\frac{(\beta_{i}^*-\beta_i')(\beta_i'-\beta_{i-1}^*)}{\beta_{i}^*-\beta_{i-1}^*}.
\end{split}
\end{equation}
\item
Transform $(p_i,L(p_i))$ $(i=j,\ldots,k+1,j\leqslant k)$ and $\beta_i=\frac{L(p_i)-L(p_{i-1})}{p_i-p_{i-1}}$ $(i=j+1,\ldots,k+1,j\leqslant k)$ to $(p_i',L(p_i)')$ and $\beta_i'$ via the following affine transformation:
\begin{equation}
    \left\{
      \begin{array}{lll}
        x'=a_{j2}(x-p_j)+b_{j2}(y-L(p_j)),& \hbox{$a_{j2}=\frac{1}{1-p_{j}} ,b_{j2}=0$,}\\
        y'=c_{j2}(x-p_j)+d_{j2}(y-L(p_j)), & \hbox{$d_{j2}=\frac{p_{j}-p_{j-1}}{p_{j}-p_{j-1}-L(p_{j})+L(p_{j-1})+p_{j-1}L(p_{j})-p_{j}L(p_{j-1})}$},\\                                         & \hbox{$c_{j2}=\frac{1-d_{j2}(1-L(p_{j}))}{1-p_{j}}$.}
      \end{array}
    \right.
    \end{equation}
Then
\begin{equation}\label{eq:Deltastar2}
\underset{\beta_i^*\in[\beta_i,\beta_{i+1}],i=j+1,\ldots,k}{\sup}\Delta^{(j,2)}=
\begin{vmatrix}
      a_{j2} & b_{j2} \\
      c_{j2} & d_{j2} \\
      \end{vmatrix}^{-1}
\underset{\beta_i^*\in[\beta_i',\beta_{i+1}'],i=j+1,\ldots,k}{\sup}\Delta^{'(j,2)}
\triangleq \begin{vmatrix}
      a_{j2} & b_{j2} \\
      c_{j2} & d_{j2} \\
      \end{vmatrix}^{-1}(\Delta^{'(j,2)})^*,
\end{equation}
where
\begin{equation}
\begin{split}
\Delta^{'(j,2)}&=
\Bigg\{(p_{j+1}'-p_j')^2(\beta_{j+1}^*-\beta_{j+1}')\frac{\beta_{j+1}'}{\beta_{j+1}^*}
+(p_{k+1}'-p_{k}')^2(\beta_{k+1}'-\beta_{k}^*)\\
&+\sum_{i=j+2}^{k}(p_{i}'-p_{i-1}')^2\frac{(\beta_{i}^*-\beta_i')(\beta_i'-\beta_{i-1}^*)}{\beta_{i}^*-\beta_{i-1}^*}\Bigg\}.
\end{split}
\end{equation}
\end{enumerate}
\end{Lemma}

\begin{figure}[!ht]
\begin{center}
  \centering
  \includegraphics[width=9cm]{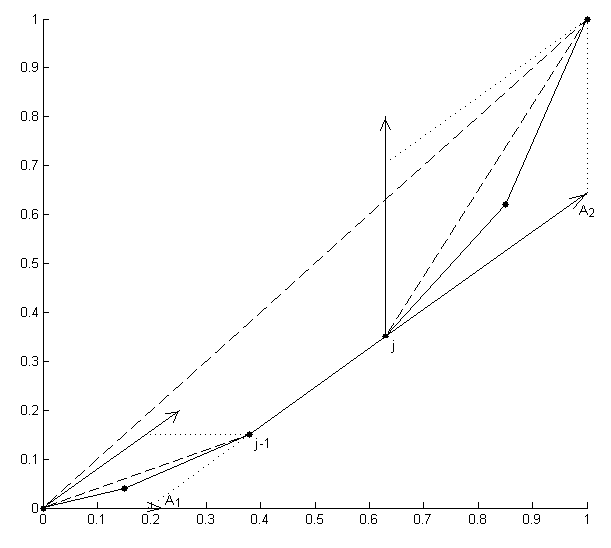}
\caption{Decomposition of $\Delta$ in Lemma~\ref{th:decompose}}\label{fig:decompose}
\end{center}
\end{figure}

\begin{Remark}[A geometric interpretation of Lemmas~\ref{th:decompose} and~\ref{th:trans}]

It is noteworthy to consider the geometric interpretation of $\Delta$ as elucidated by \cite{mehran1975bounds}: $\frac{1}{2}\Delta$ corresponds to the area enclosed between two piecewise functions. One of these functions is constructed by connecting the points $(p_i, L(p_i))$ with straight lines in sequential order, while the other is formed by the successive intersections of $l_i$ (representing the straight line passing through $(p_i, L(p_i))$ with the slope of $\beta_i^*$).

When $\beta_{j-1}^*=\beta_j^*=\beta_j$, the area $\frac{1}{2}\Delta$ is divided into two parts: $\frac{1}{2}\Delta^{(j,1)}$, and $\frac{1}{2}\Delta^{(j,2)}$ (see Figure~\ref{fig:decompose}). $\Delta^{(j,1)}$ is only related to $(p_0, L(p_0)),\ldots,(p_{j-1}, L(p_{j-1}))$, which are in a smaller coordinate system on the left of Figure~\ref{fig:decompose}. $\Delta^{(j,2)}$ is only related to $(p_j, L(p_j)),\ldots,(p_{k+1}, L(p_{k+1}))$, which are in another smaller coordinate system on the right of Figure~\ref{fig:decompose}. After affine transformation, the regions corresponding to $\frac{1}{2}\Delta^{(j,1)}$ and $\frac{1}{2}\Delta^{(j,2)}$ become the regions corresponding to $\frac{1}{2}\Delta^{'(j,1)}$ and $\frac{1}{2}\Delta^{'(j,2)}$, which have the same form as $\frac{1}{2}\Delta$. So the areas of $\frac{1}{2}\Delta^{(j,1)}$ and $\frac{1}{2}\Delta^{(j,2)}$ are $\frac{1}{2}\Delta^{'(j,1)}$ and $\frac{1}{2}\Delta^{'(j,2)}$ multiplied by the determinants of the respective affine transformations. Therefore, to maximize $\frac{1}{2}\Delta^{(j,1)}$ and $\frac{1}{2}\Delta^{(j,2)}$, we can instead maximize $\Delta^{'(j,1)}$ and $\Delta^{'(j,2)}$. 

In Figure~\ref{fig:decompose}, $A_1=(\frac{p_{j-1}L(p_j)-p_jL(p_{j-1})}{L(p_j)-L(p_{j-1})},0)$, $A_2=(1,\frac{L(p_j)-L(p_{j-1})-p_{j-1}L(p_j)+p_{j}L(p_{j-1})}{p_{j}-p_{j-1}})$. The first transformation in Lemma~\ref{th:trans} transforms $(0,0)$, $(p_{j-1},L(p_{j-1}))$ and $A_1$ to $(0,0)$, $(1,1)$ and $(1,0)$, respectively. The second transformation in Lemma~\ref{th:trans} transforms $(p_{j},L(p_j))$, $(1,1)$ and $A_2$ to $(0,0)$, $(1,1)$ and $(1,0)$, respectively.
\end{Remark}

The subsequent theorem represents the central outcome of this section, providing a key expression for $\Delta^*$ in cases where the condition described in Theorem \ref{th:unstrict2} is not satisfied. In essence, the findings in Theorem \ref{th:enumerate} complement the results in Theorem \ref{th:unstrict2}.
\begin{Theorem}\label{th:enumerate}
Define $(\Delta^{(j)})^*$ as follows:
\begin{equation}
(\Delta^{(j)})^*\triangleq\begin{vmatrix}
      a_{j1} & b_{j1} \\
      c_{j1} & d_{j1} \\
      \end{vmatrix}^{-1}(\Delta^{'(j,1)})^*
+\begin{vmatrix}
      a_{j2} & b_{j2} \\
      c_{j2} & d_{j2} \\
      \end{vmatrix}^{-1}(\Delta^{'(j,2)})^*,
\end{equation}
where $(\Delta^{'(j,1)})^*$ and $(\Delta^{'(j,2)})^*$ are given by \eqref{eq:Deltastar1} and \eqref{eq:Deltastar2}, respectively.
If \eqref{eq:conditionB} does not hold, then
\begin{equation}
\Delta^*=\underset{j=1,\ldots,k+1}{\max}(\Delta^{(j)})^*.
\end{equation}
\end{Theorem}


Theorems~\ref{th:unstrict2} and~\ref{th:enumerate} together indicate:
\begin{itemize}
\item If \eqref{eq:conditionB} holds, Theorem~\ref{th:unstrict2} implies that  $\beta_i^*=B_i$ maximizes $\Delta(\beta_1^*,\ldots, \beta_k^*)$; 
\item If \eqref{eq:conditionB} does not hold, Theorem~\ref{th:enumerate} implies that we can first evaluate $(\Delta^{'(j,1)})^*$ and $(\Delta^{'(j,2)})^*$ for all $j=1,\ldots,k+1$, then calculate $(\Delta^{(j)})^*=\begin{vmatrix}
      a_{j1} & b_{j1} \\
      c_{j1} & d_{j1} \\
      \end{vmatrix}^{-1}(\Delta^{'(j,1)})^*
+\begin{vmatrix}
      a_{j2} & b_{j2} \\
      c_{j2} & d_{j2} \\
      \end{vmatrix}^{-1}(\Delta^{'(j,2)})^*$, and finally we have $\Delta^*(\beta_1^*, \ldots, \beta_k^*)=\underset{j=1,\ldots,k+1}{\max}(\Delta^{(j)})^*$. The challenge of evaluating $(\Delta^{'(j,1)})^*$ and $(\Delta^{'(j,2)})^*$ is equivalent to evaluating $\Delta^*$ on a smaller scale. 
\end{itemize}
This implies that the actual optimal upper bound of the Gini coefficient can be solved by a recursive procedure that is summarized in Algorithm~\ref{algorithm}.

{\footnotesize
\begin{algorithm}[!ht]
    \caption{Optimal Upper Bound for Gini Coefficient Based on Grouped Data}\label{algorithm}
    \algrenewcommand\algorithmicensure{\textbf{Output:}}
        \algrenewcommand\algorithmicrequire{\textbf{Input:}}
    \begin{algorithmic}
        \Require{Number of groups $k+1$, group proportion $\{p_i\}_{i=1}^{k+1}$, group income average $\{\mu_i\}_{
        i=1}^{k+1}$}
        \Ensure{Optimal upper bound $GU$}
        \If{$k=0$} Return $\Delta^*=1$ and $GU = 1$
        \EndIf
        \State Evaluate $\mu = \sum_{i=1}^{k+1}(p_i - p_{i-1})\mu_i, \beta_i = \mu_i/\mu, L(p_i) = \sum_{j=1}^i(p_j - p_{j-1})\mu_j/\mu, i=1,\ldots, k+1.$
        \State Evaluate $z_{k+1}=1, z_{k-i} = 2p_{k-i} - z_{k-i+1}$, $i=0,\ldots, k-1$.
        \State Evaluate $w_1 = 0, w_i = 2L(p_{i-1}) - w_{i-1}, i=2,\ldots, k+1$.
        \State Evaluate for $i=1,\ldots, k$
        $$B_i = 
  \left\{\begin{array}{ll}
    \frac{w_{i+1}-w_i}{z_{i+1}-z_i}, & \hbox{when $z_{i}\neq z_{i+1};$} \\
    \infty, & \hbox{when $z_i=z_{i+1}$ and $w_i\neq w_{i+1};$} \\
    \frac{1}{2}(\beta_i+\beta_{i+1}), & \hbox{when $z_i=z_{i+1}$ and $w_i=w_{i+1}.$}
  \end{array}\right.$$
  \If{$\beta_1<B_1<\beta_2\cdots <\beta_k < B_k < \beta_{k+1}$}
  \State Return $\Delta^* = \beta_1z_1^{2}+\underset{i=1}{\overset{k}{\sum}}(\beta_{i+1}-\beta_i)(p_{i}-z_i)^2$ and $GU=1-\sum_{i=1}^{k+1}(p_i-p_{i-1})(L(p_i)+L(p_{i-1})) + \Delta^*$
  \Else
    \For{$j=1,\ldots, k+1$}
    \State Evaluate 
    \begin{equation*}
    \begin{split}
        & a_{j1}=\frac{L(p_{j})-L(p_{j-1})}{p_{j-1}L(p_{j})-p_{j}L(p_{j-1})},b_{j1}=\frac{1-a_{j1}p_{j-1}}{L(p_{j-1})}, c_{j1}=0,d_{j1}=\frac{1}{L(p_{j-1})},\\
        & a_{j2}=\frac{1}{1-p_{j}}, b_{j2}=0, c_{j2}=\frac{1-d_{j2}(1-L(p_{j}))}{1-p_{j}},\\
        & d_{j2}=\frac{p_{j}-p_{j-1}}{p_{j}-p_{j-1}-L(p_{j})+L(p_{j-1})+p_{j-1}L(p_{j})-p_{j}L(p_{j-1})}.
    \end{split}
    \end{equation*}
    \State Evaluate $(p_i',L(p_i)')$ ($i=0,\ldots, j-1$) and $(p_i',L(p_i)')$ ($i=j,\ldots, k+1$) from $(p_i,L(p_i))$ ($i=0,\ldots, j-1$) and $(p_i,L(p_i))$ ($i=j,\ldots, k+1$) respectively via the following two affine transformations:
    \begin{equation*}
    \left\{
      \begin{array}{l}
        x'=a_{j1}x+b_{j1}y,\\
        y'=c_{j1}x+d_{j1}y.
      \end{array}
    \right. \qquad     
    \left\{
      \begin{array}{l}
        x'=a_{j2}x+b_{j2}y,\\
        y'=c_{j2}x+d_{j2}y.
      \end{array}
    \right.
    \end{equation*}
    \State Evaluate $(\Delta^{'(j,1)})^*$ and $(\Delta^{'(j,2)})^*$ by inputting $\{p_i',L(p_i)'\}_{i=0}^{j-1}$ and $\{p_i',L(p_i)'\}_{i=j}^{k+1}$ to Algorithm \ref{algorithm}, respectively. Calculate $(\Delta^{(j)})^*$ by $$(\Delta^{(j)})^*=
\begin{vmatrix}
      a_{j1} & b_{j1} \\
      c_{j1} & d_{j1} \\
      \end{vmatrix}^{-1}(\Delta^{'(j,1)})^*+
\begin{vmatrix}
      a_{j2} & b_{j2} \\
      c_{j2} & d_{j2} \\
      \end{vmatrix}^{-1}(\Delta^{'(j,2)})^*.$$
    \State 
    \State Return $\Delta^*=\underset{j=1,\ldots,k+1}{\max}(\Delta^{(j)})^*$ and $GU=1-\sum_{i=1}^{k+1}(p_i-p_{i-1})(L(p_i)+L(p_{i-1})) + \Delta^*$.
    \EndFor
  \EndIf
  \end{algorithmic}
\end{algorithm}
}

Furthermore, the efficiency of this recursive algorithm can be greatly improved by employing a matrix, denoted as $M$, to store intermediate results. Specifically, let $M(i, j)$ represent the output of this algorithm when the initial data consists of the affine-transformed values $(p_i, L(p_i)),\ldots,(p_j, L(p_j))$. In cases where these values need to be used as initial data again, utilizing $M(i, j)$ obviates the need for recalculations. 


\section{Bounds for Gini Coefficient of China (2003--2008)}\label{sec:China}

\begin{figure}
    \centering
    \includegraphics[width=.9\linewidth]{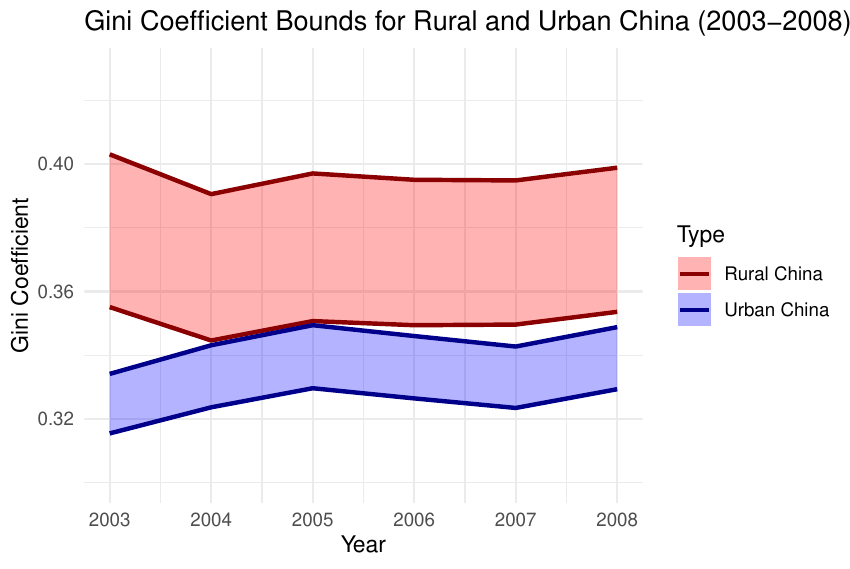}
    \caption{Gini coefficient optimal upper bounds and lower bounds for rural and urban China (2003-2008)}
    \label{fig:trend}
\end{figure}

We present the grouped data of income for Urban China and Rural China from 2003 to 2008, sourced from the {\em China Statistical Yearbook}, in Tables~\ref{tab:data1} and~\ref{tab:data2} respectively. Utilizing this data, we have computed the lower bounds of the Gini coefficients employing \eqref{eq:GL} as developed by Gastwirth \cite{gastwirth1972estimation}. Additionally, we have evaluated the optimal upper bounds using Algorithm \ref{algorithm} (referred to as Optimal Upper Bound in Tables~\ref{tab:ResultC} and~\ref{tab:ResultV}), as well as the fine upper bounds $\widetilde{GU}$ defined by \eqref{eq:tildeGU} (referred to as Fine Upper Bound in Tables~\ref{tab:ResultC} and~\ref{tab:ResultV}). The discrepancies between the optimal upper bounds and the lower bounds, termed the Width of Bounds, and the differences between the optimal upper bound and the fine upper bound, are also provided in Tables~\ref{tab:ResultC} and~\ref{tab:ResultV}. The proposed optimal upper bound provides a more refined characterization of the Gini coefficient, revealing its trend over time. See Figure \ref{fig:trend} for an illustration.

\begin{landscape}
\begin{table}
\begin{center}
\begin{threeparttable}[b]
\caption{Data of Urban China (2003--2008)}\label{tab:data1}
\begin{tabular}{ccccccccc}
  \hline
  Year & Group 1 & Group 2 & Group 3 & Group 4 & Group 5 & Group 6 & Group 7 & Group 8  \\
  \hline
  Per Capita Annual Income/Yuan& & & & & & & &\\
  \hline
  2008& 3734.35&	5754.14&	7363.28&	10195.56&	13984.23&	19254.08&	26250.10&	 43613.75 \\
  2007& 3357.91&	5058.81&	6504.60&	8900.51 &	12042.32&	16385.80&	22233.56&	 36784.51 \\
  2006& 2838.87&	4308.96&	5540.71&	7554.16 &	10269.70&	14049.17&	19068.95&	 31967.34 \\
  2005& 2495.75&	3777.53&	4885.32&	6710.58 &	9190.05 &	12603.37&	17202.93&	 28773.11 \\
  2004& 2312.50&	3428.82&	4429.05&	6024.10 &	8166.54 &	11050.89&	14970.91&	 25377.17 \\
  2003& 2098.92&	3094.93&	3970.03&	5377.25 &	7278.75 &	9763.37 &	13123.08&	 21837.32 \\
  \hline
  Population Proportion/1.0000& & & & & & & &\\
  \hline
  2008& 0.0558&	0.0568&	0.1112&	0.2114&	0.1998&	0.1896&	0.0898&	0.0857 \\
  2007& 0.0568&	0.0570&	0.1103&	0.2103&	0.1984&	0.1887&	0.0903&	0.0883 \\
  2006& 0.0565&	0.0557&	0.1088&	0.2107&	0.1988&	0.1897&	0.0914&	0.0884 \\
  2005& 0.0559&	0.0556&	0.1090&	0.2106&	0.2008&	0.1891&	0.0905&	0.0885 \\
  2004& 0.0573&	0.0556&	0.1087&	0.2085&	0.2007&	0.1904&	0.0912&	0.0876 \\
  2003& 0.0568&	0.0553&	0.1097&	0.2092&	0.2019&	0.1903&	0.0907&	0.0861 \\
  \hline
\end{tabular}
\begin{tablenotes}
{\small
   \item [1] Data source: {\em China Statistical Yearbook}. Groups 1--8 are: Poor, Lowest Income (except for Poor), Low Income, Lower Middle Income,
                            Middle Income, Upper Middle Income, High Income, Highest Income.
   \item [2] Population Proportion is the normalization of Household Proportion $\times$ Average Household Size (person).
}
\end{tablenotes}
 \end{threeparttable}
\end{center}
\end{table}
\end{landscape}

\begin{table}[!ht]
\begin{center}
\begin{threeparttable}[b]

\caption{Data of Rural China (2003--2008)}\label{tab:data2}
\begin{tabular}{cccccc}
  \hline
  Year& Group 1 & Group 2 & Group 3 & Group 4 & Group 5  \\
  \hline
  Per Capita Annual Income/Yuan& & & & &\\
  \hline
  2008& 1599.81&    2934.99&    4203.12&    5928.60&    11290.20\\
  2007& 1346.89&	2581.75&	3658.83&	5129.78&	9790.68 \\
  2006& 1182.46&	2222.03&	3148.50&	4446.59&	8474.79 \\
  2005& 1067.22&	2018.31&	2850.95&	4003.33&	7747.35 \\
  2004& 1006.87&	1841.99&	2578.49&	3607.67&	6930.65 \\
  2003& 865.90 &	1606.53&	2273.13&	3206.79&	6346.86 \\
  \hline
  Population Proportion/1.0000& & & & &\\
  \hline
  2008& 0.2263& 0.2154& 0.2029& 0.1874& 0.1680 \\
  2007& 0.2269&	0.2140&	0.2011&	0.1892&	0.1688 \\
  2006& 0.2259&	0.2131&	0.2013&	0.1889&	0.1707 \\
  2005& 0.2248&	0.2135&	0.2013&	0.1895&	0.1708 \\
  2004& 0.2255&	0.2127&	0.2010&	0.1892&	0.1716 \\
  2003& 0.2249&	0.2117&	0.2005&	0.1902&	0.1727 \\
  \hline
\end{tabular}

\begin{tablenotes}
{\small
   \item [1] Data source: {\em China Statistical Yearbook}. Group 1--5 are Low Income, Lower Middle Income, Middle Income, Upper Middle Income, High Income.
   \item [2] Population Proportion is the normalization of Household Proportion (20\%) $\times$ Average Number of Usual Residents Per Household.
}
\end{tablenotes}
\end{threeparttable}
\end{center}
\end{table}

\begin{table}[!ht]
\begin{center}
\begin{threeparttable}[b]

\caption{Bounds for Gini coefficients, Urban China (2003-2008)}\label{tab:ResultC}
\begin{tabular}{ccccccc}
  \hline
  &$2008$&$2007$&$2006$&$2005$&$2004$&$2003$\\
  \hline
  Lower Bound  & 0.3293 & 0.3234 & 0.3264 & 0.3296 & 0.3236 & 0.3154 \\
  Optimal Upper Bound & 0.3488 & 0.3427 & 0.3460 & 0.3494 & 0.3431 & 0.3341 \\
  Fine Upper Bound & 0.3605 & 0.3547 & 0.3580 & 0.3612 & 0.3543 & 0.3448 \\
  Width of Bounds      & 0.0195 & 0.0193 & 0.0196 & 0.0198 & 0.0195 & 0.0187 \\
  Fine $-$ Optimal Upper Bound & 0.0117 & 0.0120 & 0.0120 & 0.0118 & 0.0112 & 0.0107 \\
  \hline
\end{tabular}
\ \par
\caption{Bounds for Gini coefficients, Rural China (2003-2008)}\label{tab:ResultV}
\begin{tabular}{ccccccc}
  \hline
  &$2008$&$2007$&$2006$&$2005$&$2004$&$2003$\\
  \hline
  Lower Bound  & 0.3536 & 0.3496 & 0.3494 & 0.3507 & 0.3446 & 0.3551 \\
  Optimal Upper Bound & 0.3989 & 0.3949 & 0.3951 & 0.3971 & 0.3906 & 0.4031 \\
  Fine Upper Bound & 0.4059 & 0.4019 & 0.4027 & 0.4043 & 0.3985 & 0.4108 \\
  Width of Bounds    & 0.0453 & 0.0453 & 0.0457 & 0.0464 & 0.0460 & 0.0480 \\
  Fine $-$ Optimal Upper Bound & 0.0070 & 0.0070 & 0.0076 & 0.0072 & 0.0079 & 0.0077 \\
  \hline
\end{tabular}
\begin{tablenotes}
{\small
   \item [1] Lower Bounds are calculated using \eqref{eq:GL}, Optimal Upper Bounds are calculated using Algorithm \ref{algorithm}, Fine Upper Bounds are calculated using \eqref{eq:GUnotbest}, Width of Bounds are Optimal Upper Bound minus Lower Bound.
}
\end{tablenotes}
\end{threeparttable}
\end{center}
\end{table}

\section{Examples}\label{sec:example}

\begin{Example}[Counterexample for the Method in \cite{mehran1975bounds}]\label{ex:Mehran1}\end{Example}
Set $(p_1,L(p_1))=\left(\frac{3}{10},\frac{1}{10}\right)$, $(p_2,L(p_2))=\left(\frac{3}{5},\frac{4}{15}\right)$, and $(p_3,L(p_3))=\left(\frac{9}{10},\frac{8}{15}\right)$. Then $\beta_1=\frac{1}{3}$, $\beta_2=\frac{5}{9}$, $\beta_3=\frac{8}{9}$ and $\beta_4=\frac{14}{3}$. According to \eqref{eq:MehranBeta} given by \cite{mehran1975bounds}, $\beta_1^*=\frac{5}{9}$, $\beta_2^*=\frac{5}{9}$, $\beta_3^*=2$, and $\Delta=0.0617$. Our recursive algorithm  yields $\beta_1^*=\frac{5}{9},\beta_2^*=\frac{5}{9},\beta_3^*=\frac{14}{9}$, which outputs $\Delta = 0.0631 > 0.0617$. Thus, the method in \cite{mehran1975bounds} does not maximize $\Delta$ subject to the constraint~\eqref{ineq:constraint-beta}.

\begin{Example}[Counterexample for the Method in \cite{mehran1975bounds}: When $p_i=\frac{i}{k+1}$]\label{ex:Mehran2}\end{Example}
Set $(p_1,L(p_1))=\left(\frac{1}{4},\frac{1}{12}\right)$, $(p_2,L(p_2))=\left(\frac{1}{2},\frac{1}{3}\right)$ and $(p_3,L(p_3))=\left(\frac{3}{4},\frac{7}{12}\right)$. Then $\beta_1=\frac{1}{3}$, $\beta_2=1$, $\beta_3=1$ and $\beta_4=\frac{5}{3}$. According to \eqref{eq:equally} given by \cite{mehran1975bounds}, $\beta_1^*=\frac{1}{3}$, $\beta_2^*=\beta_3^*=1$, and $\Delta=0.0417$.

While using our recursive algorithm given by Section~\ref{sec:algorithm}, firstly we got
$(z_1,w_1)=(0,0),(z_2,w_2)=(\frac{1}{2},\frac{1}{6}),(z_3,w_3)=(\frac{1}{2},\frac{1}{2}),(z_4,w_4)=(1,\frac{2}{3})$,
and then $B_1=\frac{1}{3},B_2=\infty, B_3=\frac{1}{3}$.
Since $B_2>\beta_3$ does not satisfy condition~\eqref{eq:condition2}, we start calculating $(\Delta^{(j)})^*=\begin{vmatrix}
      a_{j1} & b_{j1} \\
      c_{j1} & d_{j1} \\
      \end{vmatrix}^{-1}(\Delta^{'(j,1)})^*
+\begin{vmatrix}
      a_{j2} & b_{j2} \\
      c_{j2} & d_{j2} \\
      \end{vmatrix}^{-1}(\Delta^{'(j,2)})^*,i=1,2,3,4$,
      where $(\Delta^{'(j,1)})^*$ and $(\Delta^{'(j,2)})^*$ are got using this algorithm again. Here we get $\Delta^{(1)}=\Delta|_{\beta_1^*=\frac{1}{3},\beta_2^*=1,\beta_3^*=1}=0.0417$, $\Delta^{(2)}=\Delta|_{\beta_1^*=1,\beta_2^*=1,\beta_3^*=1}=0.0556$, $\Delta^{(3)}=\Delta|_{\beta_1^*=1,\beta_2^*=1,\beta_3^*=1}=0.0556$, $\Delta^{(4)}=\Delta|_{\beta_1^*=1,\beta_2^*=1,\beta_3^*=\frac{5}{3}}=0.0139$. Therefore, $\Delta^*=\underset{i}{\max}\Delta^{(i)}=0.0556>0.0417$. So the method in \cite{mehran1975bounds}  does not maximize $\Delta$ subject to the constraint \eqref{ineq:constraint-beta}. 

\begin{Example}[Counterexample for the method in \cite{silber1990new}]\label{ex:Silber}\end{Example}
Set $(p_1,L(p_1))=\left(\frac{1}{4},\frac{1}{20}\right)$ and $(p_2,L(p_2))=\left(\frac{19}{20},\frac{3}{4}\right)$. Then $\beta_1=\frac{1}{5}$, $\beta_2=1$ and $\beta_3=5$. Using \eqref{eq:Silber} given by \cite{silber1990new}, which indicates $\beta_i^*=\frac{L(p_{i+1})-L(p_{i-1})}{p_{i+1}-p_{i-1}}$, we get $\beta_1^*=\frac{15}{19},\beta_2^*=\frac{19}{15}$, and $\Delta=0.0763$. However, using our algorithm in Section~\ref{sec:algorithm}, we get $\beta_1^*=\frac{1}{5},\beta_2^*=5$, and $\Delta= 0.3267> 0.0763$. The method in \cite{silber1990new} does not maximize $\Delta$ subject to the constraint \eqref{ineq:constraint-beta}.

\section{Discussions}\label{sec:discussion}

In this paper, we address the challenge of calculating the optimal upper bound of the Gini coefficient for grouped data without specified income brackets. We introduce an efficient, non-iterative algorithm with provable guarantees, which involves finite steps and explicit calculation formulas. To illustrate the practical utility of our approach, we present computed results for the Gini coefficients of urban and rural China from 2003 to 2008.

Our study assumes that the reported grouped income data, $\mu_i$, accurately represent the actual average income of each group. However in practice, due to sampling uncertainty, including both variance and bias, $\mu_i$ may deviate from the true group income average. An interesting direction for future research would be to develop methods for estimating lower and upper bounds for the Gini coefficient that take into account this sampling uncertainty.

\appendix{}

\section*{Appendix: Additional Proofs}

\noindent\textbf{Proof of Lemma \ref{lemma1}.} Since $f$ is convex, we have $f'_+(a) = u \leq \beta \leq f'_-(b) = v$ and for any $x\in [a, b]$,
\begin{equation*}
    f(x) \geq f(a) + (x-a)\cdot f'_+(a) ~~ \text{and} ~~ f(x) \geq f(b) - (b-x)\cdot f'_-(b).
\end{equation*}
When $v <\infty$, this means for any $a\leq x\leq b$,
\begin{equation*}
\begin{split}
    f(x) \geq & \min\left\{f(a) + (x-a)\cdot u, f(b) - (b-x)\cdot v\right\}\\
    = & \left\{\begin{array}{ll}
        f(a) + (x-a)\cdot u, & a\leq x\leq \frac{f(a)-f(b)+bv-au}{v-u}; \\
        f(b) - (b-x)\cdot v, & \frac{f(a)-f(b)+bv-au}{v-u} < x \leq b;
    \end{array}\right.
\end{split}
\end{equation*}
when $v = +\infty$, this means for any $x\geq a$,
$$f(x) \geq f(a) + (x-a) \cdot u.$$
By integrating the inequality above, we obtain the desired inequalities of Lemma \ref{lemma1}. \qed

\ \par

\noindent\textbf{Proof of Theorem~\ref{th:unstrict1}.} To prove Theorem~\ref{th:unstrict1}, we first introduce a series of notations. Define $\Delta_i, p_i^*, q_i^*$ as follows:
\begin{equation}
\begin{split}
&\Delta_1 \triangleq (p_1-p_0)^2(\beta_1^*-\beta_1)\frac{\beta_1}{\beta_1^*};\\
&\Delta_i \triangleq
\left\{
  \begin{array}{ll}
    (p_{i}-p_{i-1})^2\frac{(\beta_{i}^*-\beta_{i})(\beta_{i}-\beta_{i-1}^*)}{\beta_{i}^*-\beta_{i-1}^*}, & \hbox{$\beta_{i-1}^*\neq\beta_i^*,$}\\
    0, & \hbox{$\beta_{i-1}^*=\beta_i^*.$}
  \end{array}
\right. i=2,\ldots,k; \\
&\Delta_{k+1} \triangleq (p_{k+1}-p_{k})^2(\beta_{k+1}-\beta_{k}^*).
\end{split}
\end{equation}
Let $l_i$ be the straight line $y=\beta_i^*(x-p_i)+L(p_i), i=1,\ldots, k$, and let $l_0$ and $l_{k+1}$ be the straight lines $y=0$ and $x=1$, respectively.
Set $(p_i^*,q_i^*)$~$(i=1,\ldots,k+1)$ as the intersection of $l_{i-1}$ and $l_i$. 
Some calculating yields $$p_1^*=p_1(1-\beta_1/\beta_1^*), \quad p_i^* =\frac{p_{i}(\beta_{i}^*-\beta_i)+p_{i-1}(\beta_i-\beta_{i-1}^*)}{\beta_{i}^*-\beta_{i-1}^*}, 
 i=2,\ldots, k, \quad  p_{k+1}^*=1.$$

\begin{Lemma}\label{lemma2}
If $\{\beta_i^*\}_{i=1}^k$ satisfy
\begin{equation}\label{eq:condition2}
\beta_1< \beta_1^* < \beta_2 < \beta_2^* < \cdots < \beta_{k} < \beta_{k}^* < \beta_{k+1},
\end{equation}
then for $i=2,\ldots,k$,
\begin{equation}\label{eq:unequal2}
\Delta_i  \leqslant (\beta_i-\beta_{i-1}^*)(z-p_{i-1})^2+(\beta_{i}^*-\beta_{i})(p_{i}-z)^2,\quad \text{for all } z \in \mathbb{R},
\end{equation}
and \begin{equation}\label{eq:unequal1}
\Delta_1\leqslant \beta_1z^{2}+(\beta_1^*-\beta_1)(p_1-z)^2,\quad \text{for all } z \in \mathbb{R}.
\end{equation}
Moreover, the ``='' in \eqref{eq:unequal2} holds if and only if $z=p_i^*$, and
the ``='' in \eqref{eq:unequal1} holds if and only if $z=p_1^*$.
\end{Lemma}

\noindent\textbf{Proof of Lemma~\ref{lemma2}.} For $i=2,\ldots,k$, define $f_i(z)= \Delta_i-(\beta_i-\beta_{i-1}^*)(z-p_{i-1})^2-(\beta_{i}^*-\beta_{i})(p_{i}-z)^2$. Then
\begin{equation}
\frac{\partial f_i}{\partial z}=0 \quad \Leftrightarrow \quad z= \frac{p_{i}(\beta_{i}^*-\beta_i)+p_{i-1}(\beta_i-\beta_{i-1}^*)}{\beta_{i}^*-\beta_{i-1}^*}=p_i^*.
\end{equation}
Because $\frac{\partial ^2 f_i}{\partial z^{2}} =2(\beta_{i-1}^*-\beta_{i}^*) < 0$, $f_i(p_i^*)$ is the maximum of $f_i(z)$, which is 0. Therefore, $f_i(z)\leqslant f_i(p_i^*)=0, for\ all\ z \in R$, here ``=''holds if and only if $z=p_i^*$.\\
 Define $f_1(z)=\Delta_1-\beta_1z^2-(\beta_1^*-\beta_1)(p_1-z)^2$. Similarly, we  get $f_1(z)\leqslant f_1(p_1^*)=f_1\left(p_1\left(1-\frac{\beta_1}{\beta_1^*}\right)\right)=0, for\ all\ z \in R$, here ``=''holds if and only if $z=p_1^*$.
\qed

\ \par

According to Lemma~\ref{lemma2}, under condition \eqref{eq:condition2},
\begin{equation*}
\begin{split}
\Delta_1 &\leqslant \beta_1z_1^{2}+(\beta_1^*-\beta_1)(p_1-z_1)^2,\\
\Delta_i &\leqslant (\beta_i-\beta_{i-1}^*)(z_i-p_{i-1})^2+(\beta_{i}^*-\beta_{i})(p_{i}-z_i)^2,\quad i=2,\ldots,k.
\end{split}
\end{equation*}
Notice that $\Delta_{k+1}=(\beta_{k+1}-\beta_k^*)(p_{k+1}-p_k)^2$. So under condition \eqref{eq:condition2},\\
\begin{align*}
\Delta &=\sum_{i=1}^{k+1}\Delta_i \\
&\leqslant\beta_1z_1^{2}+(\beta_1^*-\beta_1)(p_1-z_1)^2+\sum_{i=2}^{k}(\beta_{i}^*-\beta_i)(p_{i}-z_i)^2\\
&+\sum_{i=2}^{k}(\beta_i-\beta_{i-1}^*)(z_i-p_{i-1})^2+(\beta_{k+1}-\beta_{k}^*)(p_{k+1}-p_{k})^2\\
&=\beta_1z_1^{2}+(\beta_1^*-\beta_1)(p_1-z_1)^2+\sum_{i=2}^{k}(\beta_{i}^*-\beta_i)(p_{i}-z_i)^2\\
&+\sum_{i=2}^{k}(\beta_i-\beta_{i-1}^*)(p_{i-1}-z_{i-1})^2+(\beta_{k+1}-\beta_{k}^*)(p_k-z_k)^2 \\
&=\beta_1z_1^{2}+\sum_{i=1}^{k}\{(\beta_{i+1}-\beta_{i}^*)+(\beta_{i}^*-\beta_i)\}(p_{i}-z_i)^2\\
&=\beta_1z_1^{2}+\sum_{i=1}^{k}(\beta_{i+1}-\beta_i)(p_{i}-z_i)^2.\\
\end{align*}
Therefore, the inequality in this theorem holds under condition \eqref{eq:condition2}, i.e.,  on open set $\{\beta_i<\beta_i^*< \beta_{i+1}~(i=1,\ldots,k)\}$.
Since $\Delta$ is continuous on $\{\beta_i\leqslant \beta_i^* \leqslant \beta_{i+1}~(i=1,\ldots,k)\}$ with respect to $\beta_i^*~(i=1,\ldots,k)$, the inequality in this theorem also holds on the closure of this open set, i.e., under condition \eqref{ineq:constraint-beta}. As a result, $\beta_1z_1^{2}+\overset{k}{\underset{i=1}{\sum}}(\beta_{i+1}-\beta_i)(p_{i}-z_i)^2$ is an upper bound of $\Delta$ subject to the constraint \eqref{ineq:constraint-beta}, and $\widetilde{GU}$ in \eqref{eq:tildeGU} is an upper bound of G.
\qed

\ \par

\noindent\textbf{Proof of Theorem~\ref{th:unstrict2}.} It can be tested that when \eqref{eq:conditionB} holds
and $\beta_i^*=B_i~(i=1,\ldots,k)$,
$$p_i^*=
\frac{p_{i}(\beta_{i}^*-\beta_{i})+p_{i-1}(\beta_{i}-\beta_{i-1}^*)}{\beta_{i}^*-\beta_{i-1}^*}
=z_i,\quad i=1,\ldots,k+1.$$
This makes the equalities in \eqref{eq:unequal2} and \eqref{eq:unequal1} hold, and therefore the equality in \eqref{eq:unequal} holds. Meanwhile, $\beta_i^*=B_i(i=1,\ldots,k)$ satisfies \eqref{eq:condition2}, and therefore satisfies \eqref{ineq:constraint-beta}. All of the above imply that
$\Delta^*=\beta_1z_1^{2}+\sum_{i=1}^{k}(\beta_{i+1}-\beta_i)(p_{i}-z_i)^2$, which leads to $GU=\widetilde{GU}$.
\qed

\ \par

\noindent\textbf{Proof of Lemma~\ref{th:possible}.} If $\beta_{i+1}^*=\beta_{i+1}$ and $\beta_{i-1}^*=\beta_i$ hold simultaneously, then $\Delta_i=0$ and $\Delta_{i+1}=0$, $\Delta$ does not rely on $\beta_i^*$.

 If $\beta_{i+1}^*=\beta_{i+1}$ or $\beta_{i-1}^*=\beta_i$ does not hold, then $b_i^*$ can be defined.
\begin{equation}
\begin{split}
& \frac{\partial \Delta}{\partial \beta_i^*}=0 \\
& \Leftrightarrow \beta_i^*=\frac{\beta_{i+1}^*(p_i-p_{i-1})(\beta_{i}-\beta_{i-1}^*)+\beta_{i-1}^*(p_{i+1}-p_i)(\beta_{i+1}^*-\beta_{i+1})}{(p_i-p_{i-1})(\beta_{i}-\beta_{i-1}^*)+(p_{i+1}-p_i)(\beta_{i+1}^*-\beta_{i+1})} = b_i^*
\end{split}
\end{equation}

\begin{equation}
\frac{\partial^2 \Delta}{\partial \beta_i^{*2}}=
-2\{ \frac{(p_{i}-p_{i-1})^2(\beta_{i}-\beta_{i-1}^*)^2}{(\beta_{i}^*-\beta_{i-1}^*)^3}
+\frac{(p_{i+1}-p_{i})^2(\beta_{i+1}^*-\beta_{i+1})^2}{(\beta_{i+1}^*-\beta_{i}^*)^3}\}.
\end{equation}

If $\beta_{i-1}^*$ and $\beta_{i+1}^*$ are furthermore fixed and satisfy $\beta_{i-1}\leq \beta_{i-1}^* \leq \beta_i \leq \beta_i^*\leq \beta_{i+1}$, then $\frac{\partial^2 \Delta}{\partial \beta_i^{*2}}\leqslant 0$. Therefore, the $\beta_i^*$ which maximizes $\Delta$ subject to~\eqref{ineq:constraint-beta} must satisfy
\begin{equation}
\beta_i^*=\left\{
  \begin{array}{ll}
    \beta_{i}, & \hbox{$b_i^*\leqslant \beta_{i}$}, \\
    b_i^*, & \hbox{$\beta_{i} < b_{i}^* < \beta_{i+1}$}, \\
    \beta_{i+1}, & \hbox{$b_i^*\geqslant \beta_{i+1},$}
  \end{array}
\right.
\end{equation}
which completes the proof.
\qed

\ \par

\noindent\textbf{Proof of Lemma~\ref{th:notmiddle}.} Suppose this lemma is wrong. Then with the assumption that \eqref{eq:conditionB} does not hold,
there may still exist a group of $\beta_i^* (i=1,\ldots,k)$ s.t. $\beta_i^*\in (\beta_i,\beta_{i+1})$
and at the same time $\Delta(\beta_1^*,\ldots,\beta_k^*)=\Delta^*$.
According to Lemma~\ref{th:possible}, $\beta_i^*=b_i^*$, $b_i^*$ given by \eqref{eq:bstar}.
Then it can be calculated that:
\begin{equation*}
\beta_i^*=b_i^*, for\ all\ i \quad\Leftrightarrow\quad p_{i}^*+p_{i+1}^*=2p_i, for\ all\ i \quad\Leftrightarrow\quad p_i^*=z_i, for\ all\ i.
\end{equation*}
Notice that $(p_i^*,q_i^*),(p_i,L(p_i)),(p_{i+1}^*,q_{i+1}^*)$ are all on the line of $l_i$ ($l_i$ is defined by $y=\beta_i^*(x-p_i)+L(p_i)$, $(p_i^*,q_i^*)=l_i\cap l_{i-1}$).
Thus, $q_{i}^*+q_{i+1}^*=2L(p_i), for\ all\ i$, which leads to $q_i^*=w_i, for\ all\ i$.
Since $\beta_i^* \in (\beta_i,\beta_{i+1})$ ensures $p_i^*\neq p_{i+1}^*$, i.e., $z_i\neq z_{i+1}$, we have
$$\frac{w_{i+1}-w_i}{z_{i+1}-z_i}=\frac{q_{i+1}^*-q_i^*}{p_{i+1}^*-p_i^*}=
\beta_i^* \in (\beta_i,\beta_{i+1}), for\ all\ i.$$
Namely, \eqref{eq:conditionB} holds, which contradicts the assumptions at the beginning of this proof.
As a result, the lemma holds.
\qed

\ \par

\noindent\textbf{Proof of Lemma~\ref{th:decompose}.} We will address two scenarios separately: one where $\beta_j^* =\beta_j$ and another where $\beta_j^* =\beta_{j+1}$.
\begin{itemize}
\item The case $\beta_j^*=\beta_{j}$

For $j\geqslant2$,
$$\Delta|_{\beta_j^*=\beta_{j}}=\underset{i=1}{\overset{j-1}{\sum}}\Delta_i+ \Delta_{j+1}|_{\beta_j^*=\beta_{j}}+\underset{i=j+2}{\overset{k+1}{\sum}}\Delta_i.$$
$\Delta|_{\beta_j^*=\beta_{j}}$ is non-decreasing with respect to $\beta_{j-1}^*$,
so $\beta_{j-1}^*=\beta_{j}$ makes $\Delta|_{\beta_j^*=\beta_{j}}$ the largest value under condition~\eqref{ineq:constraint-beta}, which is:
$$\Delta|_{\beta_j^*=\beta_{j},\beta_{j-1}^*=\beta_j}=\sum_{i=1}^{j-2}\Delta_i+\Delta_{j-1}|_{\beta_{j-1}^*=\beta_{j}}
+\Delta_{j+1}|_{\beta_j^*=\beta_{j}}+\sum_{i=j+2}^{k+1}\Delta_i=\Delta^{(j,1)}+\Delta^{(j,2)}.
$$
 For $j=1$, $\Delta|_{\beta_j^*=\beta_{j}}=
\Delta_{j+1}|_{\beta_j^*=\beta_{j}}+\underset{i=j+2}{\overset{k+1}{\sum}}\Delta_i=\Delta^{(j,1)}+\Delta^{(j,2)}$.\\
 To sum up, $\underset{\beta_i^*\in[\beta_i,\beta_{i+1}]}{\sup}\Delta|_{\beta_j^*=\beta_{j}}
=\underset{\beta_i^*\in[\beta_i,\beta_{i+1}],i=1,\ldots,j-2}{\sup}\Delta^{(j,1)}
+\underset{\beta_i^*\in[\beta_i,\beta_{i+1}],i=j+1,\ldots,k}{\sup}\Delta^{(j,2)}$.\\
\item The case $\beta_j^*=\beta_{j+1}$

For $j\leqslant k-1$,
$$\Delta|_{\beta_j^*=\beta_{j+1}}=\underset{i=1}{\overset{j-1}{\sum}}\Delta_i+\Delta_{j}|_{\beta_j^*=\beta_{j+1}}+\underset{i=j+2}{\overset{k+1}{\sum}}\Delta_i.$$
$\Delta|_{\beta_j^*=\beta_{j+1}}$ is nonincreasing with respect to $\beta_{j+1}^*$,
so $\beta_{j+1}^*=\beta_{j+1}$ makes $\Delta|_{\beta_j^*=\beta_{j+1}}$ the largest value under condition~\eqref{ineq:constraint-beta}, which is:
$$\Delta|_{\beta_j^*=\beta_{j+1},\beta_{j+1}^*=\beta_{j+1}}=\sum_{i=1}^{j-1}\Delta_i+\Delta_{j}|_{\beta_{j}^*=\beta_{j+1}}
+\Delta_{j+2}|_{\beta_{j+1}^*=\beta_{j+1}}+\sum_{i=j+3}^{k+1}\Delta_i=\Delta^{(j+1,1)}+\Delta^{(j+1,2)}.
$$
 For $j=k$, $\Delta|_{\beta_j^*=\beta_{j+1}}=\underset{i=1}{\overset{j-1}{\sum}}\Delta_i+
\Delta_{j}|_{\beta_j^*=\beta_{j+1}}=\Delta^{(j+1,1)}+\Delta^{(j+1,2)}$.

To sum up, $\underset{\beta_i^*\in[\beta_i,\beta_{i+1}]}{\sup}\Delta|_{\beta_j^*=\beta_{j+1}}
=\underset{\beta_i^*\in[\beta_i,\beta_{i+1}],i=1,\ldots,j-1}{\sup}\Delta^{(j+1,1)}
+\underset{\beta_i^*\in[\beta_i,\beta_{i+1}],i=j+2,\ldots,k}{\sup}\Delta^{(j+1,2)}$.
\end{itemize}
\qed

\noindent\textbf{Proof of Theorem~\ref{th:enumerate}.} According to Lemma~\ref{th:possible}, we know that even if \eqref{eq:conditionB} does not hold,
the $\beta_i^*$ that maximize $\Delta$ under condition \eqref{ineq:constraint-beta} would still be in one of the three situations listed in Lemma~\ref{th:possible}.
Meanwhile, according to Lemma~\ref{th:notmiddle}, in order to maximize $\Delta^*$,
there must exist an $s\in\{1,\ldots,k\}$ $s.t.$ $\beta_s^*$ is not in Situation (2),
i.e., $\beta_s^*=\beta_{s}$ or $\beta_s^*=\beta_{s+1}$(put Situation (3) into Situation (1)).
Then, according to Lemma~\ref{th:decompose} and Lemma~\ref{th:trans}, we get:
\begin{equation*}
\begin{split}
\Delta^*&=\underset{\beta_i^*\in[\beta_i,\beta_{i+1}],i=1,\ldots,k}{\sup}\Delta\\
&=\max\{\underset{\beta_i^*\in[\beta_i,\beta_{i+1}]}{\sup}\Delta|_{\beta_s^*=\beta_{s}}, \underset{\beta_i^*\in[\beta_i,\beta_{i+1}]}{\sup}\Delta|_{\beta_s^*=\beta_{s+1}}\}\\
&=\max\{\underset{\beta_i^*\in[\beta_i,\beta_{i+1}],i=1,\ldots,s-2}{\sup}\Delta^{(s,1)}
+\underset{\beta_i^*\in[\beta_i,\beta_{i+1}],i=s+1,\ldots,k}{\sup}\Delta^{(s,2)}, \\ & \underset{\beta_i^*\in[\beta_i,\beta_{i+1}],i=1,\ldots,s-1}{\sup}\Delta^{(s+1,1)}
+\underset{\beta_i^*\in[\beta_i,\beta_{i+1}],i=s+2,\ldots,k}{\sup}\Delta^{(s+1,2)}\}\\
&=\max\{(\Delta^{(s)})^*,(\Delta^{(s+1)})^*\}.
\end{split}
\end{equation*}
According to the definition of $\Delta^*$, for all $j=1,\ldots,k+1$, $(\Delta^{(j)})^*\leqslant\Delta^*$.
Therefore, $\max\{(\Delta^{(s)})^*,(\Delta^{(s+1)})^*\}=\underset{j=1,\ldots,k+1}{\max}(\Delta^{(j)})^*$, and $$\Delta^*=\underset{j=1,\ldots,k+1}{\max}(\Delta^{(j)})^*.$$
\qed

\bibliographystyle{plain}
\bibliography{reference}

\end{document}